\documentstyle[12pt]{article}
\newcommand\be{\begin{equation}}
\newcommand\ee{\end{equation}}
\newcommand\bea{\begin{eqnarray}}
\newcommand\eea{\end{eqnarray}}

\newcommand{\fatalpha}{{\bf \alpha \kern -0.44em \alpha}}
\newcommand{\fatsigma}{{\bf \sigma \kern -0.54em \sigma}}
\newcommand{\tpchi}{{\bf \chi \kern -0.35em \chi}}
\newcommand{\llambda}{{\bf \lambda \kern -0.45em \lambda}}

\renewcommand{\theequation}{\arabic{equation}}
\renewcommand{\theequation}{\thesection-\arabic{equation}}
\bibliography{plain}
\title{\bf Concurrence for a two-qubits mixed state consisting of three
pure states in the framework of SU(2) coherent states}
\author{ S. Salimi$^a$\thanks{Corresponding author: shsalimi@uok.ac.ir},
          A. Mohammadzade$^a$\thanks{Amir.Mohammadzade@uok.ac.ir}
          and K. Berrada$^b$\thanks{berradakamal2006@gmail.com}\\
          {$^a$\footnotesize \emph{Department of Physics, University of Kurdistan,}}\\
           {\footnotesize \emph{P.O.Box 66177-15175 , Sanandaj, Iran}.}\\
            {$^b$\footnotesize \emph{Laboratoire de Physique Th\'{e}orique URAC 13, Facult\'{e} des Sciences,}}\\
          {\footnotesize \emph{Av. Ibn Battouta, B.P. 1014, Agdal, Rabat, Morocco}.}
        }

\pagebreak

\begin{document}
\date{}
\maketitle

\begin{abstract}

A simplified expression of concurrence for two-qubit mixed state
having no more than three non-vanishing eigenvalues is obtained.
Basing on $SU(2)$ coherent states, the amount of entanglement of
two-qubit pure states is studied and conditions for entanglement
are calculated by formulating the measure in terms of some new
parameters (amplitudes of coherent states). This formalism is
generalized to the case of two-qubit mixed states using the
simplified expression of concurrence.

\section{Introduction}

$\indent$Quantum information processing essentially depends on
various quantum-mechanical phenomena, among which entanglement has
been considered as one of the most crucial features. Quantum
entanglement plays an important role in several fields of quantum
information such as quantum teleportation \cite {1,2}, quantum
cryptography \cite {5,7}, quantum dense coding \cite {8,9,10} and
quantum computation \cite {11,15,16}, etc. The fundamental question
in quantum entanglement phenomenon is: which states are entangled
and which ones are not? We can find the simple answer to this
question only in some cases. Then, the quantification and
characterization of the amount of entanglement have attracted much
attention \cite {28,29,30} in this field. For quantifying the amount
of entanglement, various measures have been proposed such as
concurrence \cite {17,18,19,20}, negativity \cite {21,22,24} and
tangle \cite {25,26,27}, etc.
\\Generally, for bipartite system pure states, the entanglement
measures are extensively accepted. However, in the case of mixed
states, the quantification of entanglement becomes more
complicated so the entanglement measures are not easy to
calculate analytically and there is no general method. The
entanglement measures of a mixed state is defined as the average
entanglement measure of an ensemble of pure states representing
the mixed state, minimized over all decompositions of the mixed
state. So the problem is to find such minimization. For various
particular cases, many minimizations can be done analytically
\cite{a,b}, but in general the problem is not solved
mathematically and it is still in early stages with many open
questions. In the case of a pair of qubits, there is a famous
formula for the bipartite concurrence and the associated
entanglement of formation which have been proposed by Wootters and
Hill \cite{42}.
\\Another concept that has vital application in quantum information
processing is the theory of coherent states for which preliminary
concepts were presented by Schr\"{o}dinger \cite {43}. Coherent
states play a crucial role in quantum physics, particularly, in
quantum optics \cite {44,45,66,67} and encoding quantum
information on continuous variables \cite {46},etc. They also
play an important role in mathematical physics \cite {49}, for
example, they are very useful in performing stationary phase
approximations to path integral \cite {50,51,52}. Ones of the
practical coherent states are SU(1,1) and SU(2) coherent states
which are widely used in entangled nonorthogonal states studying.
The entangled nonorthogonal states are very useful tools in the
quantum cryptography and quantum information processing \cite
{53}. Bosonic entangled coherent state (Glauber coherent states),
SU(1,1) and SU(2) coherent states are typical examples of
entangled nonorthogonal states.
\\In quantum statistical mechanics,
a macrostate of given system is characterized by a probability
distribution on a certain set of microstates representing the
physical properties of the system. This distribution describes
the probability of finding the system in certain microstate. In
Boltzmann's definition, entropy of the system is defined as a
measure of the number of possible microstates. The entropy grows
with this number. In this paper, we will consider a two-qubit
system described by a mixed state (macro state) defined as a
statistical mixture of three pure states representing microstates
of the system. However, the entropy will be greater than that
given in [36] and the system reaches to equilibrium and will be
more stable than that represented in \cite {56}.
\\Berrada et \textit{al} \cite {56} have used the simplified
expression of concurrence for studying the entanglement of
two-qubit mixed states having no more than two non-zero
eigenvalues (i.e., mixed states defined as statistical mixture of
two states) which play an important role in quantum information
theory using SU(2) coherent states realization. This present work
extends the research conducted by Berrada et \textit{al} \cite
{56,68} in the case of two-qubit mixed states consisting of three
pure states.
\\This paper is organized as follows: In Sect. 2 the concept of
arbitrary state of two- qubit is described and a expression for
concurrence of a two-qubit mixed state including three orthogonal
pure states is presented. Sect. 3 generalizes the findings of
sect. 2 using the formalism of SU(2) coherent states. The
conclusion and some points for further research is given in Sect.
4.
\end{abstract}

\section{Concurrence of an arbitrary state of two-qubit system}
\subsection{The case of pure state}

$\indent$In this section, we give an outline of the concurrence and
its quantity for pure and mixed states.
\\A general pure state of two-qubit system can be expressed in the
standard computational basis
 $\{|00\rangle,|01\rangle,|10\rangle|11\rangle\}$ as
\begin{equation}
|\psi\rangle = a|00\rangle + b|01\rangle
 + c|10\rangle + d|11\rangle,
\end{equation}
where $a, b, c$ and $d$ are complex numbers satisfying the
normalization condition
$$|a|^{2} + |b|^{2} + |c|^{2} +
|d|^{2}=1.$$ The concurrence for a two-qubit state $|\psi\rangle$
may be written as
\begin{equation}\label{1}
 \matrix{C(|\psi\rangle) =\mid \langle\psi|\tilde{\psi}\rangle \mid\\
 \cr\qquad\qquad = 2 \mid ad - bc \mid}
\end{equation}
where $|\tilde{\psi}\rangle =
(\sigma_{y}\otimes\sigma_{y})|\psi^{\ast}\rangle$ represents the
spin-flip plus phase flip operation. $|\psi^{\ast}\rangle$ and
$\sigma_{y}$ are the complex conjugate of $|\psi\rangle$ in the
standard basis such as
$\{|00\rangle,|01\rangle,|10\rangle,|11\rangle\}$ and pauli operator
in local basis $\{|0\rangle,|1\rangle\}$, respectively. The
concurrence is equal to $0$ for a separable state and to $1$ for a
maximally entangled states ($i.e.,
|\psi\rangle={1\over\sqrt{2}}\left(|00\rangle\pm|11\rangle\right)
\;or\; |\psi\rangle={1\over\sqrt{2}}
\left(|01\rangle\pm|01\rangle\right)$). \\The relation between
concurrence and entanglement of a pure state is given by
\begin{equation}\label{ci}
E(|\psi\rangle) = \xi(C(|\psi\rangle)),
\end{equation}
where the function  $\xi$ is defined as
\begin{equation}
\xi(C) = h\left(\frac{1+\sqrt{1-C^{2}}}{2}\right)
\end{equation}
such that
\begin{equation}\label{2}
h(x) = -x\log_{2}x - (1-x)\log_{2}(1-x)
\end{equation}
is the binary entropy function with argument related the concurrence
given by Eq. (\ref{1}). The entanglement of formation is a
monotonously increasing function of the concurrence that ranges from
$0$ for a separable state to $1$ for a maximally entangled states
(i.e.,
$|\psi\rangle={1\over\sqrt{2}}\left(|00\rangle\pm|11\rangle\right)$
or
$|\psi\rangle={1\over\sqrt{2}}\left(|01\rangle\pm|10\rangle\right)$).
Therefore, one can consider concurrence directly as a measure of the
entanglement.
\subsection{The case of mixed state}
$\indent$In the case of mixed state, the two-qubit quantum state
must be represented not by a bracket
 but a matrix called density operator and denoted by  $\rho$ in quantum mechanics. It is always to decompose
 $\rho$ into a mixture of the density operator of a set of pure states $|\psi_i\rangle$ as
\begin{equation}\label{3}
\rho = \sum_{i}p_{i}|\psi_{i}\rangle\langle\psi_{i}|,
\end{equation}
where $\{|\psi_i\rangle\}$ are distinct normalized (not necessary orthogonal) two-qubit pure states given by
\begin{equation}
|\psi_i\rangle=a_i|00\rangle + b_i|01\rangle
 + c_i|10\rangle + d_i|11\rangle
\end{equation}
and $\{p_i\}$ are the corresponding probabilities (i.e.,
$p_i\geq0$ and $\sum_ip_i=1$).
\\There is a condition for
separability and inseparability of mixed states like the pure
states which are mentioned above. A mixed state $\rho$ is said to
be separable if it can be written as a convex sum of separable
pure states, i.e., $\rho =
\sum_{i}p_{i}\rho_{i}^{(A)}\otimes\rho_{i}^{(B)}$, where
$\rho_{i}^{(A,B)}$ is the reduced density operator of qubit
($A$,$B$), respectively, given by
$\rho_{i}^{(A,B)}=Tr_{(A,B)}\left(|\Psi_{i}\rangle\langle\Psi_{i}|\right)$.
The state $\rho$ is entangled if it cannot be represented as a
mixture of a separable pure states.
\\We can define the concurrence of the mixed state $\rho$ as a
convex
 roof method which is the average concurrence of an ensemble pure states
 of the decomposition, minimized over all decomposition of
 $\rho$
\begin{equation}\label{4}
C(\rho) = \inf\sum_{i}p_{i}C(|\psi_{i}\rangle),
\end{equation}
where $C(|\Psi_{i}\rangle)$ is the concurrence of the pure state $|\Psi_{i}\rangle$ given
by Eq. (\ref{1}). According to Ref. \cite {42}, Wootters and Hill have found an explicit formula
of the concurrence defined as
\begin{equation}
C(\rho) = \max\{\lambda_{1} - \lambda_{2} - \lambda_{3} - \lambda_{4},0\}.
\end{equation}
Here $\lambda_{i}$ is the square root of eigenvalues of
$\rho(\sigma_{y}\otimes\sigma_{y})\rho^{\ast}(\sigma_{y}\otimes\sigma_{y})$
in decreasing order ($\rho^{\ast}$ denotes the complex conjugate of $\rho$).
\\In general, for a two-qubit states with no more than
two-non-zero eigenvalues which have been studied, there is an
explicit formula of the square of the concurrence \cite{56}. For
a mixed state with no more than three non vanishing eigenvalues
(i.e., state with three orthogonal pure states), Eq. (\ref{3})
can be written as
\begin{equation}
\rho = \mu_{1}|\mu_{1}\rangle\langle\mu_{1}| +
\mu_{2}|\mu_{2}\rangle\langle\mu_{2}| +
\mu_{3}|\mu_{3}\rangle\langle\mu_{3}|,
\end{equation}
where $|\mu_{1}\rangle$, $|\mu_{2}\rangle$ and $|\mu_{3}\rangle$ are three pure states given by
\begin{equation}
\matrix{|\mu_{1}\rangle = a_{1}|00\rangle + b_{1}|01\rangle +
c_{1}|10\rangle + d_{1}|11\rangle \cr |\mu_{2}\rangle =
a_{2}|00\rangle + b_{2}|01\rangle + c_{2}|10\rangle +
d_{2}|11\rangle \cr |\mu_{3}\rangle = a_{3}|00\rangle +
b_{3}|01\rangle + c_{3}|10\rangle + d_{3}|11\rangle.}
\end{equation}
So for this mixed state, Eq. (\ref{4}) is written as
\begin{equation}\label{con1}
C(\rho)=\inf(\mu_{1}C_{1} +\mu_{2}C_{2} + \mu_{3}C_{3}.
\end{equation}
In the proposed mixed state, we obtain the minimum of the
Eq.(\ref{con1}) as
\begin{equation}\label{5}
\matrix{C^{2}(\rho) = \left(\mu_{1}^{2}C_{1}^{2} + \mu_{2}^{2}C_{2}^{2} +
  \mu_{3}^{2}C_{3}^{2}\right)\cr
  + \frac{1}{2}\mu_{1}\mu_{2}\mid \textbf{C}^1 + \textbf{C}^2 -
 \textbf{C}^3 - \textbf{C}^4\mid^{2} - \frac{1}{2}\mu_{1}\mu_{2}\mid \left(\textbf{C}^1 + \textbf{C}^2 -
 \textbf{C}^3 - \textbf{C}^4\right)^{2} - 4\textbf{c}_{1}\textbf{c}_{2}\mid
 \cr+ \frac{1}{2}\mu_{1}\mu_{3}\mid \textbf{C}^1 + \textbf{C}^3 -
 \textbf{C}^2 - \textbf{C}^4\mid^{2} - \frac{1}{2}\mu_{1}\mu_{3}\mid \left(\textbf{C}^2 + \textbf{C}^3 -
 \textbf{C}^2 - \textbf{C}^4\right)^{2} - 4\textbf{c}_{1}\textbf{c}_{3}\mid
  \cr+ \frac{1}{2}\mu_{2}\mu_{3}\mid \textbf{C}^1 + \textbf{C}^4 -
 \textbf{C}^2 - \textbf{C}^3\mid^{2} - \frac{1}{2}\mu_{2}\mu_{3}\mid \left(\textbf{C}^1 + \textbf{C}^4 -
 \textbf{C}^2 - \textbf{C}^3\right)^{2} - 4\textbf{c}_{2}\textbf{c}_{3}\mid,}
 \end{equation}
where
\begin{equation}
 C_{i} =  \mid \textbf{c}_{i}\mid  =  2\mid a_{i}d_{i}-b_{i}c_{i}\mid  \qquad (i = 1,2,3)
 \end{equation}
is the concurrence of the pure state $|\mu_{i}\rangle$, and
\begin{equation}
 \matrix{C^1 =|\textbf{C}^1|= \frac{2}{3}\mid (a_{1} + a_{2} + a_{3})(d_{1} + d_{2} + d_{3}) -
 (b_{1} + b_{2} + b_{3})(c_{1} + c_{2} + c_{3}) \mid \cr C^2 = |\textbf{C}^2|=\frac{2}{3}\mid (a_{1} + a_{2} - a_{3})(d_{1} + d_{2} - d_{3}) -
 (b_{1} + b_{2} - b_{3})(c_{1} + c_{2} - c_{3}) \mid \cr C^3 = |\textbf{C}^3|=\frac{2}{3}\mid (a_{1} - a_{2} + a_{3})(d_{1} - d_{2} + d_{3}) -
 (b_{1} - b_{2} + b_{3})(c_{1} - c_{2} + c_{3}) \mid \cr C^4 = |\textbf{C}^4|=\frac{2}{3}\mid (a_{1} - a_{2} - a_{3})(d_{1} - d_{2} - d_{3}) -
 (b_{1} - b_{2} - b_{3})(c_{1} - c_{2} - c_{3}) \mid}
 \end{equation}
are the concurrences of the pure states
\begin{equation}
\matrix{|\mu^{1}\rangle = \frac{1}{\sqrt{3}}(|\mu_{1}\rangle +
 |\mu_{2}\rangle + |\mu_{3}\rangle) \cr |\mu^{2}\rangle = \frac{1}{\sqrt{3}}(|\mu_{1}\rangle +
 |\mu_{2}\rangle - |\mu_{3}\rangle) \cr |\mu^{3}\rangle = \frac{1}{\sqrt{3}}(|\mu_{1}\rangle -
 |\mu_{2}\rangle + |\mu_{3}\rangle) \cr |\mu^{4}\rangle = \frac{1}{\sqrt{3}}(|\mu_{1}\rangle -
 |\mu_{2}\rangle - |\mu_{3}\rangle)}
 \end{equation}
respectively, where $\textbf{C}^i$ and $\textbf{c}_i$ are the
complex concurrences of the pure states in Eq. (2-16) and Eq.
(2-11), respectively.  For simplicity form of Eq. (2-13),
considering the following change of variable
\begin{equation}
\matrix{C_{+} =  \mid \textbf{c}_{+} \mid =
|\textbf{C}^1+\textbf{C}^2| =
  \frac{4}{3}\mid (a_{1}+ a_{2})(d_{1}+ d_{2}) -
 (b_{1}+ b_{2})(c_{1}+ c_{2}) + a_{3}d_{3} - b_{3}c_{3} \mid \cr C_{-} = \mid \textbf{c}_{-} \mid = |\textbf{C}^3+\textbf{C}^4| =
  \frac{4}{3}\mid (a_{1}- a_{2})(d_{1}- d_{2}) -
 (b_{1}- b_{2})(c_{1}- c_{2}) + a_{3}d_{3} - b_{3}c_{3} \mid \cr C_{+}^{'} = \mid \textbf{c}_{+}^{'} \mid = |\textbf{C}^1+\textbf{C}^3| =
  \frac{4}{3}\mid (a_{1}+ a_{3})(d_{1}+ d_{3}) -
 (b_{1}+ b_{3})(c_{1}+ c_{3}) + a_{2}d_{2} - b_{2}c_{2} \mid \cr C_{-}^{'} = \mid \textbf{c}_{-}^{'} \mid = |\textbf{C}^2+\textbf{C}^4| =
  \frac{4}{3}\mid (a_{1}- a_{3})(d_{1}- d_{3}) -
 (b_{1}- b_{3})(c_{1}- c_{3}) + a_{2}d_{2} - b_{2}c_{2} \mid \cr C_{+}^{''} = \mid \textbf{c}_{+}^{''}\mid = |\textbf{C}^1+\textbf{C}^4| =
 \frac{4}{3}\mid (a_{2}+ a_{3})(d_{2}+ d_{3}) -
 (b_{2}+ b_{3})(c_{2}+ c_{3}) + a_{1}d_{1} - b_{1}c_{1} \mid \cr C_{-}^{''} = \mid \textbf{c}_{-}^{''}\mid = |\textbf{C}^2+\textbf{C}^3| =
 \frac{4}{3}\mid (a_{2}- a_{3})(d_{2}- d_{3}) -
 (b_{2}- b_{3})(c_{2}- c_{3}) + a_{1}d_{1} - b_{1}c_{1} \mid.}
 \end{equation}
Finally,  Eq. (\ref{5}) is written as
\begin{equation}\label{10}
\matrix{C^{2}(\rho) = \left(\mu_{1}^{2}C_{1}^{2} +
\mu_{2}^{2}C_{2}^{2} +
  \mu_{3}^{2}C_{3}^{2}\right)\cr
  + \frac{1}{2}\mu_{1}\mu_{2}\mid \textbf{c}_{+}-\textbf{c}_{-} \mid^{2}
  - \frac{1}{2}\mu_{1}\mu_{2}\mid (\textbf{c}_{+}-\textbf{c}_{-})^{2} - 4\textbf{c}_{1}\textbf{c}_{2}\mid
  \cr+ \frac{1}{2}\mu_{1}\mu_{3}\mid \textbf{c}_{+}^{'} - \textbf{c}_{-}^{'}\mid^{2}
  - \frac{1}{2}\mu_{1}\mu_{3}\mid (\textbf{c}_{+}^{'} - \textbf{c}_{-}^{'})^{2} - 4\textbf{c}_{1}\textbf{c}_{3}\mid
  \cr+\frac{1}{2}\mu_{2}\mu_{3}\mid \textbf{c}_{+}^{''} - \textbf{c}_{-}^{''} \mid^{2}
  -\frac{1}{2}\mu_{2}\mu_{3}\mid (\textbf{c}_{+}^{''} - \textbf{c}_{-}^{''})^{2} -
  4\textbf{c}_{2}\textbf{c}_{3}\mid.}
  \end{equation}
This equation is the square of the concurrence for a mixed state
that consists of three orthogonal pure states. Obviously, for a pure
state $\mu_1=1,\ \mu_2=0,\ \mu_3=0$ or $\mu_1=0,\ \mu_2=1,\ \mu_3=0$
or $\mu_1=0,\ \mu_2=0,\ \mu_3=1,$ our expression of concurrence
backs to the definition (\ref{1}). Also by omitting one of the pure
states of this work, we could reach the same result in  Ref. [36].
The advantages of this formula are: the concurrence of mixed state
is expressed as a function of the concurrence of the pure states and
their simple combinations, and also it can be analyzed easily.
\section{Concurrence in the language of SU(2) coherent states}
\subsection{The case of pure state}
 $\indent$The physical important of
coherent states in quantum information theory is due to the fact
that they are robust states which are widely used and applied for
studying and solving different problems in various quantum
information processing and transmission tasks, and they are easy to
generate experimentally and convenient to use. One of these states
is $SU(2)$ coherent state which is one of the most important tool
for analyzing entangled nonorthogonal states.
\\By using a phase factor, a qubit can be written as follows
\begin{equation}\label{6}
\matrix{|\theta,\varphi\rangle =
 \exp\left[-\frac{\theta}{2}\left(\sigma_{+}e^{-i\varphi}-\sigma_{-}e^{i\varphi}\right)\right]|1\rangle\\
\cr= \cos\frac{\theta}{2}|0\rangle +
 e^{i\varphi}\sin\frac{\theta}{2}|1\rangle,}
 \end{equation}
where $\sigma_{\pm} = \sigma_{x} \pm i\sigma_{y}$. Here $\sigma_{x}, \sigma_{y}$ and
$(\theta,\varphi)$ are pauli matrices and real parameters, respectively.
\\It can be proved that Eq. (\ref{6}) presents exactly an $SU(2)$
coherent state (spin coherent state) of the Klauder-Peremolov
\cite {49}.
\\The $SU(2)$ coherent state can be expressed as
\begin{equation}\label{7}
\matrix{|\gamma,j\rangle\equiv R(\gamma)|0,j\rangle
 = \exp[-\frac{1}{2}(J_{+}e^{-i\varphi}-J_{-}e^{i\varphi})]|0,j\rangle \\
 \cr=
 (1+ \mid\gamma\mid^{2})^{-j}\sum^{2j}_{n=0}\left(\matrix{2j \cr n }\right)^{\frac{1}{2}}\gamma^{n}|n,j\rangle,}
\end{equation}
\\where $R(\gamma)$ is the rotation operator, $J_{-}$ and $J_{+}$ are
the raising and lowering operators of the $su(2)$ Lie algebra,
respectively. The generators of the $su(2)$ Lie algebra, $J_{\pm}$
and $J_{z}$, satisfy the following commutation relations
\begin{equation}
\matrix{[J_{+}, J_{-}] = 2J_{z} \cr
[J_{z},J_{\pm}] = \pm J_{\pm},}
\end{equation}
and act on an irreducible unitary representation as follows
\begin{equation}
\matrix{J_{\pm}|j,m\rangle = \sqrt{(j \mp m)(j \pm m + 1)}|j,m \pm 1\rangle
 \cr J_{z}|j,m\rangle = m|j,m\rangle.}
 \end{equation}
 The $SU(2)$ coherent state can be obtained by applying successively the raising
 operator on the state $|j,-j\rangle$
\begin{equation}
|\gamma,j\rangle = \frac{1}{(1+\mid\gamma\mid^{2})^{j}}
 \sum^{j}_{m=-j}\left[\frac{(2j)!}{(j+m)!(j-m)!}\right]^{\frac{1}{2}}\gamma^{j+m}|j,m\rangle.
 \end{equation}
A change of  variable $n = j + m$ will give the form of Eq.
(\ref{7}). For a particle with spin ${1\over2}$, we get
 \begin{equation}\label{8}
 \matrix{|\gamma,\frac{1}{2}\rangle = \frac{1}{(1+\mid\gamma\mid^{2})^{\frac{1}{2}}}
 \sum^{1}_{n=0}\left(\matrix{1 \cr n }\right)\gamma^{n}|n,\frac{1}{2}\rangle\\
 \cr=
 \frac{1}{(1+\mid\gamma\mid^{2})^{\frac{1}{2}}}\left(|0,\frac{1}{2}\rangle + \gamma|1,\frac{1}{2}
 \rangle\right)}
 \end{equation}
and for $\gamma = \tan\left(\frac{\theta}{2}\right)e^{i\varphi}$, we
find that
\begin{equation}
|\gamma\rangle =
 \cos\frac{\theta}{2}|0\rangle + e^{i\varphi}\sin\frac{\theta}{2}|1\rangle
 \end{equation}
where $|0,\frac{1}{2}\rangle \equiv |0\rangle$ and
$|1,\frac{1}{2}\rangle \equiv |1\rangle$ are considered as basis
states. So, representation of a qubit using a phase factor as shown
in Eq. (\ref{6}) is equivalent to a particle with spin $\frac{1}{2}$
in construction SU(2) coherent state.
\\This result shows that the treatment and transmission of the
quantum information can be performed by using the SU(2) coherent
states.
\\Generally, a separable pure state of two-qubit system can be
written as $|\theta_{1},\varphi_{1}\rangle \otimes
|\theta_{2},\varphi_{2}\rangle.$ Considering that $
|\theta_{1},\varphi_{1}\rangle$ and
$|\theta_{1}^{'},\varphi_{1}^{'}\rangle$ are the normalized
states of the qubit $1$ and similarly for
$|\theta_{2},\varphi_{2}\rangle$ and
$|\theta_{2}^{'},\varphi_{2}^{'}\rangle$ of the qubit $2$, such
that
\begin{equation}
\langle\theta_{1},\varphi_{1}|\theta_{1}^{'},\varphi_{1}^{'}\rangle\neq0
 ,\;\;\;\;\ \langle\theta_{2},\varphi_{2}|\theta_{2}^{'},\varphi_{2}^{'}\rangle
 \neq0,
 \end{equation}
thus, the simplest extension of the arbitrary separable pure state
to an entangled pure state of  two-qubit system can be expressed by
the unnormalized state
\begin{equation}
||\psi\rangle =
 \cos\theta|\theta_{1},\varphi_{1}\rangle\otimes|\theta_{2},\varphi_{2}\rangle
 + e^{i\phi}\sin\theta|\theta_{1}^{'},
 \varphi_{1}^{'}\rangle\otimes|\theta_{2}^{'},\varphi_{2}^{'}\rangle.
 \end{equation}
Using Eq. (\ref{8}), the unnormalized state can be written as
  \begin{equation}\label{9}
  ||\psi\rangle = \cos\theta|\alpha\rangle\otimes|\beta\rangle
 + e^{i\phi}\sin\theta|\alpha^{'}\rangle\otimes|\beta^{'}\rangle
 \end{equation}
where
\begin{equation}
    \matrix{|\alpha\rangle = \frac{1}{\sqrt{(1+|\alpha|^{2})}}(|0\rangle + \alpha|1\rangle) \cr
   |\beta\rangle = \frac{1}{\sqrt{(1+|\beta|^{2})}}(|0\rangle + \beta|1\rangle) \cr
   |\alpha^{'}\rangle = \frac{1}{\sqrt{(1+|\alpha^{'}|^{2})}}(|0\rangle + \alpha^{'}|1\rangle)
   \cr |\beta^{'}\rangle = \frac{1}{\sqrt{(1+|\beta^{'}|^{2})}}(|0\rangle + \beta^{'}|1\rangle)}
\end{equation}
are respectively the states for each qubit.
\\So, Eq. (\ref{9}) becomes
\begin{equation}
\matrix{||\psi\rangle = \frac{\cos\theta}{\sqrt{(1 +
  \mid \alpha \mid^{2})(1 + \mid \beta \mid^{2}))}}(|00\rangle + \beta|01\rangle + \alpha|10\rangle +
  \alpha\beta|11\rangle)\cr+\frac{e^{i\phi}\sin\theta}{\sqrt{(1 +
  \mid \alpha^{'} \mid^{2})(1 + \mid \beta^{'} \mid^{2}))}}(|00\rangle + \beta^{'}|01\rangle +
  \alpha^{'}|10\rangle + \alpha^{'}\beta^{'}|11\rangle)
 \cr \hspace{-2.4cm}=  a|00\rangle + b|01\rangle
 + c|10\rangle + d|11\rangle},
 \end{equation}
where
 \begin{equation}
 \matrix{a = \lambda + \gamma \cr b = \beta\lambda + \beta^{'}\gamma
 \cr c = \alpha\lambda + \alpha^{'}\gamma \cr d = \alpha\beta\lambda + \alpha^{'}\beta^{'}\gamma}
 \end{equation}
with
\begin{equation}
\lambda = \frac{\cos\theta}{\sqrt{(1 +
  \mid \alpha \mid^{2})(1 + \mid \beta \mid^{2})}},\quad \gamma = \frac{e^{i\phi}\sin\theta}{\sqrt{(1 +
  \mid \alpha^{'} \mid^{2})(1 + \mid \beta^{'} \mid^{2})}}.
\end{equation}
Finally, the normalized pure state can be expressed in standard
computational basis $\{|00\rangle,|01\rangle,$
$|10\rangle,|11\rangle\}$ as
\begin{equation}\label{10}
|\psi\rangle = \frac{1}{\sqrt{N}}(a|00\rangle + b|01\rangle
 + c|10\rangle + d|11\rangle),
 \end{equation}
with
\begin{equation}
 N = \langle \psi|\psi \rangle = |a|^{2} + |b|^{2} + |c|^{2} +
 |d|^{2}.
 \end{equation}
\\The concurrence of the state (\ref{10}) is given in terms of the
amplitudes of coherent states as
\begin{equation}
C(|\psi\rangle) = \mid \langle\psi|\tilde{\psi}\rangle \mid =
 2\left|\frac{\lambda\gamma}{N}
(\alpha - \alpha^{'})(\beta - \beta^{'})\right|.
\end{equation}
The minimum of the concurrence ($C(|\psi\rangle)=0$, i.e., the state
is separable) is attained in either of the following situations:
$\alpha=\alpha^{'}$ or $\beta=\beta^{'}$ or $\lambda=0$ or
$\gamma=0$. Furthermore, its maximum is satisfied when
$C(|\psi\rangle)=1$ which corresponds to maximally entangled states.

\subsection{The case of mixed state}
 $\indent$The realization that entanglement is a resource for a number of useful tasks in quantum information
is led to a tremendous interest in its properties, quantification,
and in method by which it can be produced. In this way, we use a
helpful method basing on spin coherent states for quantifying the
entanglement of two-qubit mixed  consisting of three pure states.
\\$\indent$Let us consider  a class of mixed states
 defined as a statistical mixture of three pure states of two-qubit system
\begin{equation}
\rho = \sum_{i}p_{i}|\psi_{i}\rangle\langle\psi_{i}|\qquad (i =
1,2,3),\end{equation} where
\begin{equation}
|\psi_{i}\rangle = \frac{1}{\sqrt{N_{i}}}(a_{i}|00\rangle
+ b_{i}|01\rangle
 + c_{i}|10\rangle + d_{i}|11\rangle)
\end{equation}
represents the pure states of two-qubit system, with
\begin{equation}
\matrix{a_{i} = \lambda_{i} + \gamma_{i} \cr
b_{i} = \beta_{i}\lambda_{i} + \beta^{'}_{i}\gamma_{i}
 \cr c_{i} = \alpha_{i}\lambda_{i} + \alpha^{'}_{i}\gamma_{i} \cr d_{i} =
 \alpha_{i}\beta_{i}\lambda_{i} + \alpha^{'}_{i}\beta^{'}_{i}\gamma_{i}}
 \end{equation}
and
\begin{equation} N_{i} = \langle \psi_{i}|\psi_{i} \rangle =
|a_{i}|^{2} + |b_{i}|^{2} + |c_{i}|^{2} + |d_{i}|^{2}.
\end{equation}
\\The expression of the concurrence of the mixed state with three
orthogonal states given by  Eq. (2-18) can be directly
generalized to the case of mixed state with three nonorthogonal
states \cite{38}. So, in our case we find that
\begin{equation}
 \matrix{C^{2}(\rho) = \left(p_{1}^{2}C_{1}^{2}
+ p_{2}^{2}C_{2}^{2} +
  p_{3}^{2}C_{3}^{2}\right)\cr
  + \frac{1}{2}p_{1}p_{2}\mid \textbf{c}_{+}-\textbf{c}_{-} \mid^{2}
  - \frac{1}{2}p_{1}p_{2}\mid (\textbf{c}_{+}-\textbf{c}_{-})^{2} - 4\textbf{c}_{1}\textbf{c}_{2}\mid
  \cr+ \frac{1}{2}p_{1}p_{3}\mid \textbf{c}_{+}^{'} - \textbf{c}_{-}^{'}\mid^{2}
  - \frac{1}{2}p_{1}p_{3}\mid (\textbf{c}_{+}^{'} - \textbf{c}_{-}^{'})^{2} - 4\textbf{c}_{1}\textbf{c}_{3}\mid
  \cr+\frac{1}{2}p_{2}p_{3}\mid \textbf{c}_{+}^{''} - \textbf{c}_{-}^{''} \mid^{2}
  -\frac{1}{2}p_{2}p_{3}\mid (\textbf{c}_{+}^{''} - \textbf{c}_{-}^{''})^{2} -
  4\textbf{c}_{2}\textbf{c}_{3}\mid,}
  \end{equation}
where
\begin{equation}
 C_{i} =  \mid \textbf{c}_{i}\mid  = 2\left|\frac{\lambda_{i}\gamma_{i}}{N_{i}}
(\alpha_{i} - \alpha^{'}_{i})(\beta_{i} -
\beta^{'}_{i})\right|\end{equation} is the concurrence of the pure
state $|\psi_{i}\rangle$,
\begin{equation}
 \matrix{C_{\pm} = \mid \textbf{c}_{\pm} \mid =
\frac{4}{3}\left|
\frac{\lambda_{1}\gamma_{1}}{N_{1}}(\alpha_{1}-\alpha_{1}^{'})(\beta_{1}-\beta_{1}^{'})
+
\frac{\lambda_{2}\gamma_{2}}{N_{2}}(\alpha_{2}-\alpha_{2}^{'})(\beta_{2}-\beta_{2}^{'})
+
\frac{\lambda_{3}\gamma_{3}}{N_{3}}(\alpha_{3}-\alpha_{3}^{'})(\beta_{3}-\beta_{3}^{'})\right.
\cr\left.\pm\frac{1}{\sqrt{N_{1}N_{2}}}(\lambda_{1}\lambda_{2}(\alpha_{1}-\alpha_{2})(\beta_{1}-\beta_{2})
+
\lambda_{1}\gamma_{2}(\alpha_{1}-\alpha_{2}^{'})(\beta_{1}-\beta_{2}^{'})
+
\lambda_{2}\gamma_{1}(\alpha_{1}^{'}-\alpha_{2})(\beta_{1}^{'}-\beta_{2})\right.
\cr\left.+
\gamma_{1}\gamma_{2}(\alpha_{1}^{'}-\alpha_{2}^{'})(\beta_{1}^{'}-\beta_{2}^{'}))\right|}
\end{equation}
\begin{equation}
\matrix{C_{\pm}^{'} = \mid \textbf{c}_{\pm}^{'} \mid =
\frac{4}{3}\left| \frac{\lambda_{1}\gamma_{1}}{N_{1}}(\alpha_{1}-\alpha_{1}^{'})(\beta_{1}-\beta_{1}^{'})
+ \frac{\lambda_{2}\gamma_{2}}{N_{2}}(\alpha_{2}-\alpha_{2}^{'})(\beta_{2}-\beta_{2}^{'})
+ \frac{\lambda_{3}\gamma_{3}}{N_{3}}(\alpha_{3}-\alpha_{3}^{'})(\beta_{3}-\beta_{3}^{'})\right.
\cr\left.\pm\frac{1}{\sqrt{N_{1}N_{3}}}(\lambda_{1}\lambda_{3}(\alpha_{1}-\alpha_{3})(\beta_{1}-\beta_{3})
+ \lambda_{1}\gamma_{3}(\alpha_{1}-\alpha_{3}^{'})(\beta_{1}-\beta_{3}^{'})
+ \lambda_{3}\gamma_{1}(\alpha_{1}^{'}-\alpha_{3})(\beta_{1}^{'}-\beta_{3})\right.
\cr\left.+ \gamma_{1}\gamma_{3}(\alpha_{1}^{'}-\alpha_{3}^{'})(\beta_{1}^{'}-\beta_{3}^{'}))\right|}
\end{equation}
and
  \begin{equation}
  \matrix{C_{\pm}^{''} = \mid \textbf{c}_{\pm}^{''} \mid =
\frac{4}{3}\left|
\frac{\lambda_{1}\gamma_{1}}{N_{1}}(\alpha_{1}-\alpha_{1}^{'})(\beta_{1}-\beta_{1}^{'})
+
\frac{\lambda_{2}\gamma_{2}}{N_{2}}(\alpha_{2}-\alpha_{2}^{'})(\beta_{2}-\beta_{2}^{'})
+
\frac{\lambda_{3}\gamma_{3}}{N_{3}}(\alpha_{3}-\alpha_{3}^{'})(\beta_{3}-\beta_{3}^{'})\right.
\cr\left.\pm\frac{1}{\sqrt{N_{2}N_{3}}}(\lambda_{2}\lambda_{3}(\alpha_{2}-\alpha_{3})(\beta_{2}-\beta_{3})
+
\lambda_{2}\gamma_{3}(\alpha_{2}-\alpha_{3}^{'})(\beta_{2}-\beta_{3}^{'})
+
\lambda_{3}\gamma_{2}(\alpha_{2}^{'}-\alpha_{3})(\beta_{2}^{'}-\beta_{3})\right.
\cr\left.+
\gamma_{2}\gamma_{3}(\alpha_{2}^{'}-\alpha_{3}^{'})(\beta_{2}^{'}-\beta_{3}^{'}))\right|.}
\end{equation}
\\The simplified expression of concurrence of the mixed state
reveals some general important features:
\\{\bf{A.}} The concurrence has an upper and lower bound expressed
as
 \begin{equation}
 (p_{1}C_{1} - p_{2}C_{2} - p_{3}C_{3})^{2}\leq C^{2}(\rho)
 \leq(p_{1}C_{1} + p_{2}C_{2} + p_{3}C_{3})^{2},
 \end{equation}
where
\begin{equation}
\matrix{(p_{1}C_{1} - p_{2}C_{2} - p_{3}C_{3})^{2} =
4(p_{1}|\frac{\lambda_{1}\gamma_{1}}{N_{1}} (\alpha_{1} -
\alpha^{'}_{1})(\beta_{1} - \beta^{'}_{1})| \cr -
p_{2}|\frac{\lambda_{2}\gamma_{2}}{N_{2}} (\alpha_{2} -
\alpha^{'}_{2})(\beta_{2} - \beta^{'}_{2})| \cr -
p_{3}|\frac{\lambda_{3}\gamma_{3}}{N_{3}} (\alpha_{3} -
\alpha^{'}_{3})(\beta_{3} - \beta^{'}_{3})|)^{2}}
\end{equation}
and
\begin{equation}
\matrix{(p_{1}C_{1} + p_{2}C_{2} + p_{3}C_{3})^{2} =
4(p_{1}|\frac{\lambda_{1}\gamma_{1}}{N_{1}} (\alpha_{1} -
\alpha^{'}_{1})(\beta_{1} - \beta^{'}_{1})| \cr +
p_{2}|\frac{\lambda_{2}\gamma_{2}}{N_{2}} (\alpha_{2} -
\alpha^{'}_{2})(\beta_{2} - \beta^{'}_{2})| \cr +
p_{3}|\frac{\lambda_{3}\gamma_{3}}{N_{3}} (\alpha_{3} -
\alpha^{'}_{3})(\beta_{3} - \beta^{'}_{3})|)^{2}}
\end{equation}
are the lower and upper bounds of concurrence, respectively.
\\{\bf{B.}} If the three pure states consisting the mixed state
have only real components (i.e., $a_i,b_i,c_i$ and $d_i$ are real
numbers), then:
\\- for
\begin{equation}
\matrix{0 \leq 4\textbf{c}_{1}\textbf{c}_{2}
\leq (\textbf{c}_{+} - \textbf{c}_{-})^{2}\cr 0 \leq
4\textbf{c}_{1}\textbf{c}_{3} \leq (\textbf{c}_{+}^{'} -
\textbf{c}_{-}^{'})^{2}\cr 0 \leq
4\textbf{c}_{2}\textbf{c}_{3} \leq (\textbf{c}_{+}^{''}
- \textbf{c}_{-}^{''})^{2},}
\end{equation}
the concurrence reaches the upper bound
\begin{equation}
C^{2}(\rho) = 4\left(p_{1}\left|\frac{\lambda_{1}\gamma_{1}}{N_{1}}
(\alpha_{1} - \alpha^{'}_{1})(\beta_{1} - \beta^{'}_{1})\right|  +
p_{2}\left|\frac{\lambda_{2}\gamma_{2}}{N_{2}} (\alpha_{2} -
\alpha^{'}_{2})(\beta_{2} - \beta^{'}_{2})\right|  +
p_{3}\left|\frac{\lambda_{3}\gamma_{3}}{N_{3}} (\alpha_{3} -
\alpha^{'}_{3})(\beta_{3} - \beta^{'}_{3})\right|\right)^{2};
\end{equation}
\\- for
\begin{equation}
\matrix{4\textbf{c}_{1}\textbf{c}_{2} \geq (\textbf{c}_{+} -
\textbf{c}_{-})^{2}\geq0\cr 4\textbf{c}_{1}\textbf{c}_{3}\geq
(\textbf{c}_{+}^{'} - \textbf{c}_{-}^{'})^{2}\geq0 \cr
4\textbf{c}_{2}\textbf{c}_{3}\geq (\textbf{c}_{+}^{''} -
\textbf{c}_{-}^{''})^{2}\geq0,}
\end{equation}
the concurrence is as follows
\begin{equation}
\matrix{ C^{2}(\rho) = 4\left(p_{1}\left|\frac{\lambda_{1}\gamma_{1}}{N_{1}}
(\alpha_{1} - \alpha^{'}_{1})(\beta_{1} - \beta^{'}_{1})\right|  -  p_{2}\left|\frac{\lambda_{2}\gamma_{2}}{N_{2}}
(\alpha_{2} - \alpha^{'}_{2})(\beta_{2} - \beta^{'}_{2})\right|  -  p_{3}\left|\frac{\lambda_{3}\gamma_{3}}{N_{3}}
(\alpha_{3} - \alpha^{'}_{3})(\beta_{3} - \beta^{'}_{3})\right|\right)^{2}\cr
+L_{\alpha_1,\alpha_2,\beta_1,\beta_2}^{\alpha_1^{'},\alpha_2^{'},\beta_1^{'},\beta_2^{'}} +
M_{\alpha_1,\alpha_3,\beta_1,\beta_3}^{\alpha_1^{'},\alpha_3^{'},\beta_1^{'},\beta_3^{'}}+
N_{\alpha_2,\alpha_3,\beta_2,\beta_3}^{\alpha_2^{'},\alpha_3^{'},\beta_2^{'},\beta_3^{'}} }
\end{equation}
where
\begin{equation}
\matrix{L_{\alpha_1,\alpha_2,\beta_1,\beta_2}^{\alpha_1^{'},\alpha_2^{'},\beta_1^{'},\beta_2^{'}}=
{64p_1p_2\over9N_1N_2}\left(\lambda_{1}\lambda_{2}(\alpha_{1}-\alpha_{2})(\beta_{1}-\beta_{2})
+
\lambda_{1}\gamma_{2}(\alpha_{1}-\alpha_{2}^{'})(\beta_{1}-\beta_{2}^{'})
+
\lambda_{2}\gamma_{1}(\alpha_{1}^{'}-\alpha_{2})(\beta_{1}^{'}-\beta_{2})\right.
\cr\left.+
\gamma_{1}\gamma_{2}(\alpha_{1}^{'}-\alpha_{2}^{'})(\beta_{1}^{'}-\beta_{2}^{'})\right)^{1\over2}}
\end{equation}
\begin{equation}
\matrix{M_{\alpha_1,\alpha_3,\beta_1,\beta_3}^{\alpha_1^{'},\alpha_3^{'},\beta_1^{'},\beta_3^{'}}=
{64p_1p_3\over9N_1N_3}\left(\lambda_{1}\lambda_{3}(\alpha_{1}-\alpha_{3})(\beta_{1}-\beta_{3})
+ \lambda_{1}\gamma_{3}(\alpha_{1}-\alpha_{3}^{'})(\beta_{1}-\beta_{3}^{'})
+ \lambda_{3}\gamma_{1}(\alpha_{1}^{'}-\alpha_{3})(\beta_{1}^{'}-\beta_{3})\right.
\cr\left.+ \gamma_{1}\gamma_{3}(\alpha_{1}^{'}-\alpha_{3}^{'})(\beta_{1}^{'}-\beta_{3}^{'})\right)^{1\over2}}
\end{equation}
\begin{equation}
\matrix{N_{\alpha_2,\alpha_3,\beta_2,\beta_3}^{\alpha_2^{'},\alpha_3^{'},\beta_2^{'},\beta_3^{'}}=
{64p_2p_3\over9N_2N_3}\left(\lambda_{2}\lambda_{3}(\alpha_{2}-\alpha_{3})(\beta_{2}-\beta_{3})
+
\lambda_{2}\gamma_{3}(\alpha_{2}-\alpha_{3}^{'})(\beta_{2}-\beta_{3}^{'})
+
\lambda_{3}\gamma_{2}(\alpha_{2}^{'}-\alpha_{3})(\beta_{2}^{'}-\beta_{3})\right.
\cr\left.+
\gamma_{2}\gamma_{3}(\alpha_{2}^{'}-\alpha_{3}^{'})(\beta_{2}^{'}-\beta_{3}^{'})\right)^{1\over2};}
\end{equation}
\\- for
\begin{equation}
\matrix{\textbf{c}_{1}\textbf{c}_{2}\leq 0
 \cr \textbf{c}_{1}\textbf{c}_{3}\leq 0   \cr  \textbf{c}_{2}\textbf{c}_{3}\leq0,}
 \end{equation}
the concurrence is equivalent to lower bound
\begin{equation}
 C^{2}(\rho) = 4\left(p_{1}\left|\frac{\lambda_{1}\gamma_{1}}{N_{1}}
(\alpha_{1} - \alpha^{'}_{1})(\beta_{1} - \beta^{'}_{1})\right|  -  p_{2}\left|\frac{\lambda_{2}\gamma_{2}}{N_{2}}
(\alpha_{2} - \alpha^{'}_{2})(\beta_{2} - \beta^{'}_{2})\right|  -  p_{3}\left|\frac{\lambda_{3}\gamma_{3}}{N_{3}}
(\alpha_{3} - \alpha^{'}_{3})(\beta_{3} - \beta^{'}_{3})\right|\right)^{2}.
\end{equation}
\\{\bf{C.}} When
 \begin{equation}
 \matrix{\textbf{c}_{+} =\textbf{c}_{-} \cr \textbf{c}_{+}^{'} =\textbf{c}_{-}^{'} \cr
 \textbf{c}_{+}^{''} =\textbf{c}_{-}^{''},}
 \end{equation}
the concurrence reaches again the lower bound
 \begin{equation}
 C^{2}(\rho) = 4\left(p_{1}\left|\frac{\lambda_{1}\gamma_{1}}{N_{1}}
(\alpha_{1} - \alpha^{'}_{1})(\beta_{1} - \beta^{'}_{1})\right|  -  p_{2}\left|\frac{\lambda_{2}\gamma_{2}}{N_{2}}
(\alpha_{2} - \alpha^{'}_{2})(\beta_{2} - \beta^{'}_{2})\right|  -  p_{3}\left|\frac{\lambda_{3}\gamma_{3}}{N_{3}}
(\alpha_{3} - \alpha^{'}_{3})(\beta_{3} - \beta^{'}_{3})\right|\right)^{2}.
\end{equation}
\\We remark that we can obtain the concurrence for many other
cases.
\\{\bf{D.}} We take the case where two pure states of mixed state
are separable (i.e, $C_{1} = C_{2} = 0 $, $C_{1} = C_{3} = 0 $ or
$ C_{2} = C_{3} = 0 $). Then, the concurrence of the mixed state
becomes as
\begin{equation}
C^{2}(\rho) = 4p_{i}^{2}\left|\frac{\lambda_{i}\gamma_{i}}{N_{i}}
(\alpha_{i}-\alpha_{i}^{'})(\beta_{i}-\beta_{i}^{'})\right|^{2}
\end{equation}
($i=1$ for $C_2=C_3=0;$ $i=2$ for $C_1=C_3=0$ and $i=3$ for
$C_1=C_2=0$).
\\We notice that from the result obtained in section (3.1), the
pure states $|\psi_i\rangle$  ($i=1,2,3$) are separable when
$\alpha_i=\alpha_i^{'}$ or $\beta_i=\beta_i^{'}$ or $\lambda_i=0$
or $\gamma_i=0.$
\\The above equation shows  that the pure state $|\psi_1\rangle$
(respectively $|\psi_2\rangle$ or $|\psi_3\rangle$) and its
probability $p_1$ (respectively $p_2$ or $p_3$) contain the vital
information about the entanglement of two-qubit mixed state.
\\Without loss of generality, we consider the simple case where
$ \alpha_{i}=\beta_{i}$ and $\alpha^{'}_{i}=\beta^{'}_{i}$, then
the concurrence is simplified as \cite{59}
\begin{equation}
\matrix{C^{2}(\rho) =
 p^{2}_{i}\frac{(\alpha_{i}-\alpha_{i}^{'})^{4}}
 {(2\alpha^{2}_{i}\alpha^{'2}_{i} + \alpha^{'2}_{i} + 2\alpha_{i}\alpha_{i}^{'}
 + \alpha^{2}_{i} + 2)^{2}}\\
 \cr= \left(\frac{p_{i}}{1+2X_{i}}\right)^{2}}
 \end{equation}
where
 \begin{equation}
 X_{i} =
 \left(\frac{\alpha_{i}\alpha_{i}^{'}+1}{\alpha_{i}-\alpha_{i}^{'}}\right)^{2}\in [0,\infty[.
 \end{equation}
\\Now we discuss two important limit cases:
\\{\bf{1.}} $X_{i}=0$
($\alpha_{i} = \frac{-1}{\alpha_{i}^{'}}$), i.e., the state
$|\psi_i\rangle$ is a Bell state, the two-qubit state
$$\rho=p_1|\psi_1\rangle\langle\psi_1|+p_2|\psi_2\rangle\langle\psi_2|+p_3|\psi_3\rangle\langle\psi_3|$$
is a statistical mixture of a Bell state and two separable pure
states which corresponds to $C^{2}(\rho) = p^{2}_{i}$. These states
represent an important class of mixed states which are widely used
and applied in different quantum information processing and
transmission tasks and have some applications.
\\{\bf{2.}} $X_{i}\rightarrow \infty$ $(\alpha_{i} =
\alpha_{i}^{'})$, i.e., the state $|\psi_i\rangle$ is separable,
the state
$$\rho=p_1|\psi_1\rangle\langle\psi_1|+p_2|\psi_2\rangle\langle\psi_2|+p_3|\psi_3\rangle\langle\psi_3|$$
is a classical mixture of three separable pure states which
corresponds to separable mixed state $C(\rho)=0.$
\\An important case is merit notifying: the completely mixed state
for which the density operator is simply $\rho = {I\over d}$ where d
is dimension of the Hilbert space (in our case $\rho ={I\over4}$),
this corresponds to $p_{1} = p_{2} = p_{3} = \frac{1}{3}$. Plotting
$C^{2}(\rho)$ as a function of X and P and also $C^{2}(\rho)$ as a
function of $\alpha$ and $\alpha^{'}$ (Fig1. (a) and (b)), we see
that the maximal value of the concurrence $C(\rho)$ is $\frac{1}{3}$
rather than $1$ as for the pure state. This concept can be explained
by the fact that in a completely mixed state, the concurrence is
equally shared by the subsystems.
\\By comparing these plots with those indicated  in Ref \cite
{56}, we notice that square of concurrence is reduced from $0.25$
to $0.11$ (or in other words) by increasing the number of
two-qubit pure states from two to three.

\section{Conclusion}
$\indent$In this paper, we have studied the entanglement of
two-qubit states, our approach was to write the measure in terms
of the amplitudes of $SU(2)$ coherent states. As a measure of
entanglement, we have used the concurrence. We  expressed it as a
function of the amplitudes of coherent states and we have given
the sufficient conditions for the minimal and maximal of
entanglement in the case of two-qubit nonorthogonal pure states.
By determining a simplified expression of concurrence for a
two-qubit mixed state having no more than three non-zero
eigenvalues, we have generalized  the formalism of two-qubit
nonorthogonal pure states to the case of a class of mixed states.
However, we have studied the behavior of the square of the
two-qubit mixed state concurrence where their conditions were
depending on both of the amplitudes and corresponding
probabilities.
\\By studying a simple case, we found that the
concurrence of the mixed state cannot be higher than the
probability of one of the qubits. Furthermore, for the
completely  mixed state it cannot exceed one third.
\\In this way, it is shown that the $SU(2)$
coherent states are useful elements
 to determine and measure the entanglement of two-qubit states and their use is not
only of the theoretical purpose but also of some practical
importance having in mind their experimental accessibility \cite
{39}. \\The two-qubit nonorthogonal states are expected to have
more applications in quantum information theory. Throughout the
paper we have only considered the bipartite entanglement. The
more difficult task is to quantify the genuine multipartite
entanglement. In this context, We intend to use and generalize
the result to the case of various qubits and consider possible
applications in quantum information.

 \vspace{1cm}
\setcounter{section}{0}
 \setcounter{equation}{0}
 \renewcommand{\theequation}{A-\arabic{equation}}


\newpage
{\bf Figure Captions}

{\bf Fig1:} (a) Schema of $C^{2}(\rho)$ as a function of X and P.
(b) Schema of $C^{2}(\rho)$ as a function of $\alpha$ and
$\alpha^{'}$ for $P_{1} = P_{2} = P_{3} = \frac{1}{3}$,
  $-5 \leq \alpha \leq 5$ and $-5 \leq \alpha^{'} \leq 5$.


\begin{thebibliography}{99}
\bibitem{1}
C. H. Bennett, G. Brassard, C. Cr´epeau, R. Jozsa, A. Peres and W.
K. Wootters, Phys. Rev. Lett. 70, 1895 (1993).

\bibitem{2}
D. Bouwmeester, J. W. Pan, K. Mattle, M. Eibl, H. Weinfurter and  A.
Zeilinger, Nature 390, 575 (1997).


\bibitem{5}
C. H. Bennet, G. Brassard, Quantum cryptography: Public key
distribution and coin tossing, Proceedings of IEEE International
Conference on Computers, Systems and Signal Processing, 175 (1984).


\bibitem{7}
A. K. Ekert. Phys. Rev. Lett. 67, 661 (1991).


\bibitem{8}
C. H. Bennett and S. J. Wiesner. Phys. Rev. Lett. 69, 2881 (1992).

\bibitem{9}
S.L. Braunstein and  H.J. Kimble, Phys. Rev. A 61, 042302 (2000).

\bibitem{10}
K. Mattle, H. Weinfurter, P. G. Kwait and A. Zeilinger, Phys. Rev.
Lett. 76, 4656 (1996).


\bibitem{11}
M. A. Nielsen  and I. L. Chuang , Quantum computation and
information,Cambridge university press (2000).


\bibitem{15}
A. Barenco, D. Deutsch, A. Ekert and R. Jozsa, Phys. Rev. Lett. 74,
4083 (1995).

\bibitem{16}
D. Deutsch, Proc. R. Soc. Lond. A 425, 73 (1985).

\bibitem{28}F. Mintert, M. Ku\'{s} and A. Buchleitner, Phys. Rev. Lett.
95, 260502 (2005).

\bibitem{29}F. Mintert and A. Buchleitner, Phys. Rev. Lett. 98,
140505 (2007).

\bibitem{30}L. Aolita, A. Buchleitner and F. Mintert, Phys. Rev. A
78, 022308 (2008).

 \bibitem{17}
C.H. Bennett, H.J. Bernstein, S. Popescu and B. Schumacher, Phys.
Rev. A 53, 2046 (1996).

 \bibitem{18}
P. Rungta, V. Bu¡zek, C. M. Caves, M. Hillery and G. J. Milburn,
Phys. Rev. A 64, 042315 (2001).

\bibitem{19}
L. M. Kuang and L. Zhou, Phys. Rev. A 68, 043606 (2003).

\bibitem{20}A. Uhlmann, Phys. Rev. A 62, 032307 (2000).

 \bibitem{21}A. Peres, Phys. Rev. Lett. 77, 1413 (1996).

\bibitem{22}M. Horodecki, P. Horodecki and R. Horodecki, Phys. Lett. A 223, 1 (1996).



\bibitem{24}G. Vidal and R. F. Werner, Phys. Rev. A 65, 032314, (2002).


\bibitem{25}V. Coffman, J. Kundu and W. K. Wootters, Phys. Rev.
A 61, 052306 (2000).

\bibitem{26}
A. Wong and N. Christensen, Phys. Rev. A 63, 044301 (2001).

\bibitem{27}
T. J. Osborne and F. Verstraete, Phys. Rev. Lett 96, 220503, (2006).

\bibitem{a}B. M. Terhal and K. G. H. Vollbrecht, Phys. Rev. Lett. A
85, 2625 (2000).

\bibitem{b}P. Rungta and C. M. Caves, Phys. Rev. A 67, 012307
(2003).


\bibitem{42}S. Hill and W. K. Wootters, Phys. Rev. Lett. 78, 5022 (1997).


 \bibitem{43}E. Schr\"{o}dinger, Naturwissenschafter. 14, 664 (1926).

 \bibitem{44}
J. R. Klauder and Bo-S. Skagerstam: Coherent States, World
Scientific, Singapore (1985).

\bibitem{45}
L. Mandel and E. Wolf : Optical Coherence and Quantum Optics,
Cambridge University Press (1995).


\bibitem{46}
S. Lloyd and S. L. Braunstein, Phys. Rev. Lett. 82, 1784 (1999).


\bibitem{49}A. Perelomov : Generalized Coherent States and Their Applications, Springer–
Verlag (1986).

\bibitem{50}K. Funahashi, T. Kashiwa, S. Sakoda and K. Fujii, J. Math. Phys. 36,3232 (1995).


\bibitem{51}K. Funahashi, T. Kashiwa, S. Sakoda and K. Fujii, J.
Math. Phys. 36, 4590 (1995).


\bibitem{52}K. Fujii, T. Kashiwa and S. Sakoda, J. Math. Phys. 37, 567 (1996).

\bibitem{53}P. D\"{o}m\"{o}t\"{o}r, M. G. Benedict, Phys. Lett. A 372, 3792 (2008).



\bibitem{56}K. Berrada, A. Chafik, H. Eleuch, Y. Hassouni, Quantum Inf Process. 9, 13 (2010).


\bibitem{59}K. Berrada, A. Chafik, H. Eleuch, Y. Hassouni, Int. Jou. Mod. Phys B. 23, 2021
(2009).
\bibitem{38}Y. M. Di, B. L. Hu, D. D. Liu, et al.: Acta. Phys. Sin.
55, 3869 (2006).
\bibitem{39}J. Katriel and A. I. Solomon,
Phys. Rev. A 49, 5149 (1994).
\bibitem{66}H. Eleuch, R. Bennaceur, J. Opt B 6, 189 (2004).
\bibitem{67}T. C. Ralph, Rep. Prog. Phys. 69, 853 (2006).
\bibitem{68}K. Berrada, M. El Baz, H. Eleuch, Y. Hassouni, Int. Jou. Mod. Phys C. 21,
291 (2010).
\end{thebibliography}
\end{document}